\documentclass[journal, twoside, web]{ieeecolor}
\usepackage{lcsys}
\usepackage{times}
\usepackage{multicol, multirow, threeparttable}
\usepackage[bookmarks=true]{hyperref} %
\usepackage{amsmath,amssymb,amsfonts}
\usepackage[pdftex]{graphicx}   
\usepackage{url}
\usepackage{xcolor}
\usepackage{caption}
\usepackage{verbatim}
\usepackage{multirow, tabularx, booktabs}
\renewcommand{\thetable}{\arabic{table}}
\usepackage{subcaption}
\usepackage{balance} %
\usepackage{algorithm}
\usepackage[noend]{algpseudocode}  %
\usepackage{lipsum}
\usepackage[%
    style=numeric-comp,
    sorting=none,
    natbib=true,  %
    backend=biber,
    doi=false,
    url=false,
    giveninits=true,
    maxcitenames=2, %
    minbibnames=3, %
    maxbibnames=4, %
    hyperref
]{biblatex}

\newbibmacro{string+doiurl}[1]{
    \iffieldundef{doi}{
        \iffieldundef{url}{#1}{\href{\thefield{url}}{#1} }
    }{\href{https://dx.doi.org/\thefield{doi}}{#1}}
}

\DeclareFieldFormat[article,incollection,techreport,inproceedings,book,misc]%
    {title}%
    {\usebibmacro{string+doiurl}{\mkbibquote{#1}}}
\renewbibmacro{in:}{}
\AtEveryBibitem{%
    \clearfield{isbn}%
    \clearfield{issn}
}

\definecolor{darkgreen}{RGB}{29, 177, 2}
\definecolor{porange}{RGB}{231, 117, 0}  %
\definecolor{nvgreen}{HTML}{76B900}  %

\algrenewcommand\algorithmicindent{10pt}
\algrenewcommand\algorithmiccomment[1]{{\hfill$\triangleright$ #1}}
\algnewcommand{\LineComment}[1]{\Statex {// #1}}

\newcommand{\such}{{\mathop{\text{s.t.}}}}
\DeclareMathOperator*{\argmax}{{\mathop{\mathrm{argmax}}}}
\DeclareMathOperator*{\argmin}{{\mathop{\mathrm{argmin}}}}

\newcommand{\reals}{\mathbb{R}}

\newcommand{\statedim}{{n}}
\newcommand{\ctrldim}{{m}}

\newcommand{\state}{{x}}
\newcommand{\ctrl}{{u}}

\newcommand{\stateImag}[2]{{\state_{#1 | #2}}}
\newcommand{\ctrlImag}[2]{{\ctrl_{#1 | #2}}}

\newcommand{\traj}{{\mathbf{x}}}
\newcommand{\csig}{{\mathbf{u}}}

\newcommand{\xSet}{{\mathcal{X}}}

\newcommand{\cSet}{{\mathcal{U}}}

\newcommand{\csigSet}{{\mathbb{U}}}

\newcommand{\dyn}{{f}} %

\newcommand{\thorizon}{{T}}

\newcommand{\tdisc}{{t}} %
\newcommand{\tdiscaux}{{\tau}}
\newcommand{\tdiscplan}{{k}}
\newcommand{\tdiscauxaux}{{s}}

\newcommand{\khorizon}{{H}}

\newcommand{\C}{{\mathcal{C}}}

\newcommand{\outcomeFunc}[3]{J\left( #2, #3, #1 \right)}

\newcommand{\valFunc}{{V}}

\newcommand{\policy}{{\pi}}

\newcommand{\consFunc}{{g}}
\newcommand{\targFunc}{{\ell}}

\newcommand{\target}{{\mathcal{T}}}

\newcommand{\failure}{{\mathcal{F}}}

\newcommand{\raHoriFunc}[3]{{\mathcal{RA}\left( #1, #2, #3 \right)}}

\newcommand{\safeSet}{{\Omega}}
\newcommand{\safeSetOpt}{{\safeSet^*}}

\newcommand{\shield}{{\text{sf}}}
\newcommand{\stopp}{{\text{stop}}}
\newcommand{\task}{{\text{p}}}
\newcommand{\backup}{{\text{b}}}
\newcommand{\policyShield}{{\policy^\shield}}
\newcommand{\policyStop}{{\policy^\stopp}}
\newcommand{\policyTask}{{\policy^\task}}
\newcommand{\policyBackup}{{\policy^\backup}}
\newcommand{\policyBackupOpt}{{\policy^{\backup*}}}
\newcommand{\policyInv}{{\policy^\target}}
\newcommand{\ctrlTask}{{\ctrl^\task}}
\newcommand{\ctrlBackup}{{\ctrl^\backup}}

\newcommand{\policyBackupSig}{{\boldsymbol{\policy}^\backup}}
\newcommand{\policyBackupIdx}[1]{{\policy^\backup_{#1}}}

\newcommand{\stateErr}[1]{{\delta \state_{#1 | \tdisc}}}
\newcommand{\ctrlErr}[1]{{\delta \ctrl_{#1}}}

\newcommand{\stateNom}[1]{{\overline{\state}_{#1|\tdisc}}}

\newcommand{\ctrlNom}[1]{{\overline{\ctrl}_{#1}}}

\newcommand{\fut}[1]{{\dyn_{\ctrl_{#1}}}}
\newcommand{\trajNom}{{\overline{\traj}}}
\newcommand{\csigNom}{{\overline{\csig}}}

\newcommand{\openloop}{{k_\tdisc}}
\newcommand{\closedloop}{{K_\tdisc}}
\newcommand{\olsig}{{\mathbf{k}}}
\newcommand{\clsig}{{\mathbf{K}}}

\newcommand{\cbfquad}{P}
\newcommand{\cbfaff}{p}
\newcommand{\cbfconst}{c}

\newcommand{\yawdeviation}{\delta \theta}
\newcommand{\optparam}{\Phi}

\newcommand{\vxx}[1]{{Z_{#1}}}
\newcommand{\vx}[1]{{z_{#1}}}

\newcommand{\barrier}{{B}}
\newcommand{\dynScale}{{\lambda}}

\newtheorem{theorem}{Theorem}

\newtheorem{lemma}{Lemma}

\usepackage[normalem]{ulem}
\newcommand{\out}[1]{}
\newboolean{showrevision}
\setboolean{showrevision}{false}
\newcommand{\new}[1]{\ifthenelse{\boolean{showrevision}}{{#1}}{#1}}
\newcommand{\newcaption}[1]{\ifthenelse{\boolean{showrevision}}{{#1}}{\caption{#1}}}

\newboolean{finalrevision}
\setboolean{finalrevision}{false}
\newcommand{\final}[1]{\ifthenelse{\boolean{finalrevision}}{\textcolor{porange}{#1}}{#1}}

\newcolumntype{C}[1]{>{\centering\arraybackslash} m{#1}}

\begin{document}

\title{
    Fast, Smooth, and Safe: Implicit Control Barrier Functions through Reach-Avoid Differential Dynamic Programming
}

\author{
Athindran Ramesh Kumar,
Kai-Chieh Hsu,
Peter J. Ramadge,
and
Jaime F. Fisac
\thanks{All authors are with the Department of Electrical and Computer Engineering, Princeton University, New Jersey, USA. email:
        {\tt\small arkumar@princeton.edu}}
}

\maketitle
\thispagestyle{empty}

\pagestyle{empty}

\begin{abstract}
Safety is a central requirement for autonomous system operation across domains.
Hamilton-Jacobi (HJ) reachability analysis can be used
to construct ``least-restrictive'' safety filters that result in infrequent, but often extreme, control overrides.
In contrast, control barrier function (CBF) methods apply smooth control corrections to guard the system against an often conservative safety boundary.
This paper provides an online scheme to construct an implicit CBF through HJ reach-avoid differential dynamic programming in a receding-horizon framework,
enabling smooth safety filtering with infinite-time safety guarantees.
Simulations with the Dubins car and 5D bicycle dynamics demonstrate the scheme's ability to preserve safety smoothly without the conservativeness of handcrafted CBFs.
\end{abstract}


\section{Introduction}

\IEEEPARstart{T}{hrough} improved sensors, computation, learning, and decision-making techniques, autonomous robotic systems are becoming increasingly capable.
However, deploying these systems in safety-critical environments requires robust fail-safe methods to ensure the avoidance of
catastrophic \textit{failure states}.
To enforce safe behavior, various modern techniques rely on finding a \textit{safe set} $\safeSet$,
that is, a subset of
non-failure states
such that from each state in $\safeSet$ there exists a control strategy that will keep the state in $\safeSet,$ i.e., $\safeSet$ is controlled-invariant. 

Hamilton-Jacobi (HJ) reachability analysis provides a constructive approach to obtain the \emph{maximal} controlled-invariant safe set $\safeSetOpt$ \citep{fisac2015reach, bansal2017hamilton}. Since there generally exist no closed-form solutions, a state-space discretization is often used to compute the value function and control policy using numerical dynamic programming (DP). These solvers  \citep{mitchell2008flexible} scale poorly with state space dimension $n$ and are impractical for $n\geq 5$. Several approximate methods \citep{chen2018decomposition, hsu2021safety, fisac2019bridging, landry2018reach, fridovichkeil2021approximate} have been proposed to overcome this issue. The safety controls obtained from reachability solvers are subsequently used in a ``least-restrictive'' (LR) framework where the safety control is only applied if the system attempts to cross the zero-level set of the HJ value function. While attractive in theory, this minimally invasive scheme can yield abrupt ``panic’’ behaviors in practice, as it waits until the last chance before applying the safety control. 
\out{
\begin{figure}    \centering
    \includegraphics[width=0.85\columnwidth]{figures_march/cover.pdf}
    \caption{CBF-DDP modifies the task policy's action using a CBF-type constraint to obtain the filtered $\policyShield$.
    For the reach-avoid objective, we define a safety margin function $\consFunc$, the distance to the failure set at each state, and a target margin function $\targFunc$, the shortest distance to the failure set along a counterfactual stopping path from each state.}
    \vspace{-5pt}
    \label{fig:cover_fig}
    \vspace{-3pt}
\end{figure}
}
Control barrier functions (CBF) \citep{ames2017cbf, agrawal2017discrete} offer an alternative approach to guarantee safety.
Given a known safe set $\Omega$, these functions are designed to guarantee its invariance working in conjunction with a quadratic program (CBF-QP).
The CBF-QP modifies the control well before the boundary of~$\Omega$, and a constraint on the derivative of the CBF prevents rapid movement towards the boundary, resulting in smoother trajectories than the LR approach. However, control barrier functions have two drawbacks. First, they are
typically handcrafted for
conservative invariant sets $\Omega\subset\Omega^*$. Second, the CBF-QP can become infeasible in the presence of tight actuation constraints, endangering safety; a backup policy is then needed to ensure persistent feasibility~\cite{chen2021backup}. 
In contrast, the invariant sets obtained via HJ reachability analysis are maximal and guarantee feasibility, but the LR controls can be extreme. 

To bridge both frameworks, \cite{choi2021robust} introduces a control-barrier value function (CBVF) derived using HJ reachability theory. The CBVF is computed offline using numerical dynamic programming.
The warm-start approach of~\cite{tonkens2022refining} alleviates the high computational burden by using a learned or handcrafted candidate CBF as initialization and iteratively refining it through numerical HJ analysis, asymptotically converging to 
a valid infinite-time CBF.
\new{However, numerical methods generally scale poorly to systems 
beyond 6 continuous state dimensions and are challenging for real-time applications.}

Here, we propose a provably safe online scheme,
which constructs an implicit CBF at each control cycle through reach-avoid differential dynamic programming (DDP).
\new{Our CBF-DDP scheme enables the \textbf{fast} runtime computation of control solutions that vary \textbf{smoothly} along system trajectories and are guaranteed to persistently enforce \textbf{safety} at all future times.}
\out{To our knowledge, this work provides the first general online constructive approach for CBFs leveraging HJ safety analysis.}%
Our race car simulations with a 5-D state space demonstrate smoother safety filtering than LR while overcoming the conservativeness of handcrafted CBFs.

\section{Preliminaries} \label{sec:background}
Let $\cSet\subseteq \reals^{\ctrldim}$ be a non-empty compact set and  
consider a discrete-time 
system with state $\state \in \xSet \subseteq \reals^{\statedim}$ and control input $\ctrl \in \cSet $, evolving as $ \state_{\tdisc+1} = \dyn(\state_\tdisc, \ctrl_\tdisc)$.
Let $[\tdisc, \khorizon] := \{ \tdisc, \tdisc+1, \cdots, \khorizon\}$ and
$\csigSet_{\tdisc:\khorizon}$ denote the 
set of 
input sequences 
$\csig_{\tdisc:\khorizon} := ( \ctrl_\tdisc, \cdots, \ctrl_{\khorizon}).$
Let $\traj\left( \tdiscaux; \new{\state_0}, \policy \right)$, $\tdiscaux \geq0$,
denote a state trajectory starting from $\new{\state_0}$ under  
control policy $\policy \colon \xSet \to \cSet$ and dynamics $\dyn$.
\out{We sometimes abbreviate $\traj\left( \tdiscaux; \new{\state_0}, \policy \right)$ as $\state_\tdiscaux$.}
For an open set $\failure \subseteq \xSet$ of failure states,
there always exists a Lipschitz-continuous 
function 
$\consFunc \colon \xSet \to \reals$ such that $\consFunc(\state) < 0 \iff \state \in \failure$
(a signed distance function). 

\subsection{Controlled-invariant Set}
A (closed) set $\safeSet \subset \xSet$ is 
\textit{controlled-invariant} under policy $\policyBackup$ if for all 
$\state_0 \in \safeSet$ and all $\tdiscaux \geq 0$, $\traj(\tdiscaux; \state_0, \policyBackup) \in \safeSet$.
The goal of safety analysis is to find a controlled-invariant set $\safeSet$, \new{and its corresponding~$\policyBackup$,}
such that 
${\safeSet \cap \failure = \emptyset}$.
In this context, $\safeSet$ is called the 
\textit{safe set}.
Additionally, to reduce conservativeness,
we desire $\safeSet$ to 
be as large as possible
(ideally, close to the maximal safe set).
Once such a safe set is constructed, a \textit{least-restrictive safety filter} $\policyShield$ 
can be 
applied to any \new{performance-driven control policy~$\policyTask \colon \xSet \to \cSet$,
enforcing the safety-oriented control~$\policyBackup(\state)$
whenever using~$\policyTask(\state)$}
would lead to leaving the safe set $\safeSet$:
\begin{IEEEeqnarray}{c}
 \policyShield(\state) := \left\{
    \begin{array}{ll}
        \policyTask(\state), & \dyn \left( \state, \policyTask(\state) \right) \in \safeSet \\
        \policyBackup(\state), & \text{otherwise.}
    \end{array}
\right. 
\label{eq:least_restrictive_ctrl}
\end{IEEEeqnarray}

\subsection{Control Barrier Functions}\label{sec:cbf}
Control barrier functions (CBFs) provide Lyapunov-like conditions to guarantee the forward invariance of a (safe) set and have been applied in various safety-critical control problems~\citep{ames2017cbf}.
\citet{agrawal2017discrete} extend the CBF initially developed for the continuous-time dynamic systems to the discrete-time domain.
Formally, they consider a continuously differentiable function $\barrier \colon \xSet \to \reals$
whose zero superlevel set is the safe set $\safeSet \subseteq \xSet$,
i.e., $\safeSet = \{ \state \in \xSet \mid \barrier(\state) \geq 0\}$.
A function $\barrier$ is a discrete-time exponential CBF for dynamics~$\dyn$ if,
\new{for some $\gamma \in (0, 1]$ and for each $\state\in\safeSet$, there exists a control $\ctrl\in\cSet$ such that}
$\barrier \big( \dyn(\state, \ctrl) \big) \geq \gamma \barrier(\state)$~\citep{agrawal2017discrete}.
In other words, such control input $\ctrl$ keeps the state trajectory inside~$\safeSet$ and ensures \new{that its boundary} %
can only be approached asymptotically.
We can incorporate this CBF constraint into an online trajectory optimization as
\begin{subequations}
\begin{IEEEeqnarray}{rcc}
    \policyShield (\state) = \ & \argmin_{\ctrl \in \cSet} & \| \ctrl - \policyTask(\state) \|_2^2 \\
    & \such & \barrier \big( \dyn(\state, \ctrl) \big) \geq \gamma \barrier(\state). \label{eq:disc_cbf_cstr}
\end{IEEEeqnarray}
\label{eq:cbf}%
\end{subequations}
If dynamics are control-affine and the CBF is quadratic, \eqref{eq:cbf} can be solved with a quadratically constrained quadratic program (QCQP) \citep{agrawal2017discrete}.
This enables real-time computation even for high-dimensional dynamics, making it attractive for many robotics applications.
However, finding a valid CBF for general dynamic systems is difficult.
Although there are successes of heuristically constructed CBFs, 
these do not recover the maximal safe set and may be overly conservative~\citep{choi2021robust}.
Further, the safety guarantee is contingent on the feasibility of the quadratic program, which could be lost in the presence of tight actuation limits or other control constraints.

\subsection{Hamilton-Jacobi Reachability Analysis}\label{sec:hj_reachability}
Hamilton-Jacobi (HJ) reachability analysis defines an optimal control problem to find the maximal safe set (or \textit{viability kernel}) $\safeSetOpt$.
The value function ${\valFunc \colon \xSet \times [0, \khorizon] \to \reals}$, $\valFunc(\new{\state_\tdisc}, \tdisc) := \max_{\policyBackup} \min_{\tdiscaux \in [\tdisc, \khorizon]} \consFunc(\traj(\tdiscaux-\tdisc; \state_\tdisc, \policyBackup))$ encodes the minimum safety margin that the safety policy $\policyBackup$ can maintain over all times up to a planning horizon $\khorizon$.
The value function is then computed by finding the solution of the finite-horizon terminal-value problem with safety Bellman equation~\citep{fisac2019bridging}
\begin{subequations}
\new{
\begin{IEEEeqnarray}{rcl}
    \valFunc(\state_\tdisc, \tdisc) & \ := \ & 
        \min\Big\{
            \consFunc(\state_\tdisc), \max_{\ctrl_\tdisc \in \cSet} \valFunc\big(\dyn(\state_\tdisc, \ctrl_\tdisc), \tdisc+1\big)
        \Big\},~ \\
    \valFunc(\state_\khorizon, \khorizon) & := & \consFunc(\state_\khorizon).
\end{IEEEeqnarray}
}%
\label{eq:hjb_reachability}%
\end{subequations}
Note that this value function only allows finite-horizon safety. 
So in practice, HJ reachability analysis usually considers a sufficiently long horizon to guarantee infinite-horizon safety, i.e., $\valFunc^\infty(\state) := \lim_{\khorizon \to \infty} \max_{\policyBackup} \min_{\tdiscaux \in [0, \khorizon]} \consFunc(\new{\traj(\tdiscaux; \state, \policyBackup)}) 
$.
\new{Numerical} level-set methods have been developed to compute the HJ value function, and require discretizing the state space into \new{a finite grid}; thus, computation scales exponentially with the dimensionality of the state space, \new{becoming practically prohibitive beyond 6 continuous state variables}~\citep{bui2022optimizeddp}.
If this value function can be computed, we can retrieve the viability kernel by $\safeSetOpt := \{\state \in \xSet \mid \valFunc^\infty(\state) \geq 0 \}$ and the safety policy $\policyBackupOpt(\state) = \argmax_{\ctrl \in \cSet} \valFunc^\infty(\dyn(\state, \ctrl))$.

Additionally, wherever this value function is differentiable, it is a valid CBF; thus, HJ reachability analysis provides a principled way to construct a CBF.
Due to the \out{maximum}\new{minimum} operator in \eqref{eq:hjb_reachability}, $\valFunc^\infty \left( \dyn(\state, \policyBackupOpt(\state)) \right) \geq \valFunc^\infty(\state)$, the safety policy~$\policyBackupOpt$ always prevents the safety value from decreasing.
Thus, this HJ value function is often paired with the least-restrictive control law in \eqref{eq:least_restrictive_ctrl}, only overriding the proposed control input near the boundary of the safe set $\safeSetOpt$~\citep{bansal2017hamilton}.
However, this least-restrictive scheme can result in jerky control switches and is susceptible to numerical issues near the safe set boundary. In contrast, CBF-constrained optimization \eqref{eq:cbf} provides a smoother transition at the cost of additional restrictions to the control input away from the safe set boundary.

\subsection{Model Predictive Control}\label{sec:mpc}
Model predictive control (MPC)~\citep{mayne2000constrained} verifies the system's safety on the fly instead of relying on a pre-computed value function or CBF.
In the context of safety filters, MPC can be used to optimize a control sequence such that the first control is the closest to the task-oriented control but satisfies all the constraints below
\begin{subequations}
\begin{IEEEeqnarray}{cl}
     \min_{\ctrl_{\tdiscplan \mid \tdisc}, \state_{\tdiscplan \mid \tdisc}} & \left\|
        \policyTask (\state_\tdisc) - \ctrl_{0 \mid \tdisc}
    \right\| \\
    \such & \forall \tdiscplan \in [0, \khorizon-1] \colon \nonumber \\
    & \state_{0 \mid \tdisc} = \state_\tdisc, \ \state_{\tdiscplan+1 \mid \tdisc} = \dyn \left( \state_{\tdiscplan \mid \tdisc}, \ctrl_{\tdiscplan \mid \tdisc} \right), \label{eq:mpc_dyn_cstr} \\
    & \stateImag{\tdiscplan}{\tdisc} \not\in \failure, %
    \ \ctrlImag{\tdiscplan}{\tdisc} \in \cSet, \label{eq:mpc_ctrl_state_cstr} \\
    & \stateImag{\khorizon}{\tdisc} \in \target \label{eq:mpc_ter_cstr}.
\end{IEEEeqnarray}
\label{eq:mpc}%
\end{subequations}
In addition to the dynamics constraint~\eqref{eq:mpc_dyn_cstr} and state and control constraints~\eqref{eq:mpc_ctrl_state_cstr}, MPC adds a terminal constraint~\eqref{eq:mpc_ter_cstr} confining the final state to a target $\target \subset \xSet$, \new{rendered controlled-invariant by a known} $\policyInv$.
After the optimization, only the first control will be executed, i.e., $\policyShield (\state_\tdisc) = \ctrl_{0 | \tdisc}$.
At the next time step $\tdisc+1$, \eqref{eq:mpc} is \new{solved} again with a \new{shifted} time horizon.
Importantly, constraint~\eqref{eq:mpc_ter_cstr} ensures recursive feasibility, as we can repeat the control sequence $(\ctrl_{1:\khorizon-1 \mid \tdisc}, \policyInv(\state_{\khorizon \mid \tdisc}))$ at $\tdisc+1$.
This recursive feasibility in turn guarantees \new{all-time} safety.

\subsection{Reach-Avoid Analysis}
MPC requires the system's state to be inside \new{a known} controlled-invariant set exactly at the end of the planning horizon.
However, we can loosen this \new{requirement} by considering a reach-avoid problem, which \new{finds} a control policy to guide the system's state into a (closed) target set at any time within the planning horizon without breaching the constraints at prior times.
Specifically, the \textit{reach-avoid} set for horizon $\khorizon$ is
\begin{IEEEeqnarray}{c}
    \raHoriFunc{\target}{\failure}{\khorizon} := \big\{ \state \in \xSet \mid
        \exists \policyBackup, \tdisc \in [0, \khorizon], \nonumber \\
    \traj\left(\tdisc; \state, \policyBackup \right) \in \target \wedge \forall \tdiscaux \in [0, \tdisc], \traj\left(\tdiscaux; \state, \policyBackup \right) \notin \failure 
    \big\}. \label{eq:reach_avoid}
\end{IEEEeqnarray}
Introducing an additional Lipschitz-continuous target margin function $\targFunc \colon \xSet \to \reals$ such that $\targFunc(\state) \geq 0 \iff \state \in \target$,
we can follow an analogous formulation to~\ref{sec:hj_reachability}, defining
the reach-avoid objective along horizon $[\tdisc, \khorizon]$ for arbitrary~${\tdisc \in [0, \khorizon]}$:
\begin{align}
    & \outcomeFunc{\tdisc}{\state_\tdisc}{\policyBackup}
    := \max_{\tdiscaux \in [\tdisc, \khorizon]} \min \Big\{
        \targFunc \big( \state_\tdiscaux \big),
        \min_{\tdiscauxaux \in [\tdisc, \tdiscaux]} \consFunc \big( \state_\tdiscauxaux \big)
    \Big\} \label{eq:outcome}
\end{align}
\new{with $\state_\tdiscaux = \traj (\tdiscaux-\tdisc; \state_\tdisc, \policyBackup)$}.
The associated value function \new{$\valFunc (\state_\tdisc, \tdisc) :=\max_{\policyBackup} \outcomeFunc{\tdisc}{\state_\tdisc}{\policyBackup}$} satisfies the reach-avoid Bellman equation~\citep[Appendix A]{hsu2021safety}:
\begin{IEEEeqnarray}{ll}
\label{eq:hjb_reachavoid}
\valFunc 
(\state_\tdisc, \tdisc) = \min \bigg\{& \consFunc(\state_\tdisc), \\
& 
\max \Big\{
            \targFunc( \state_\tdisc ),
            \max_{\ctrl_\tdisc \in \cSet} \valFunc\big(\dyn(\state_\tdisc, \ctrl_\tdisc), \tdisc+1\big)
        \Big\} \!\!\bigg\} \nonumber
    , 
\end{IEEEeqnarray}
with terminal condition 
$\valFunc (\state_\khorizon, \khorizon) = \min \left\{ \consFunc(\state_\khorizon), \targFunc(\state_\khorizon) \right\}$.
The reach-avoid set~\eqref{eq:reach_avoid} is characterized by the sign of the value function at $\tdisc=0$, as $\valFunc(\state, 0) \geq 0 \iff \state \in \raHoriFunc{\target}{\failure}{\khorizon}$.

Recent work addresses the \emph{curse of dimensionality} (the exponential scaling of computation) by approximate methods such as state-space decomposition~\cite{chen2018decomposition}, deep reinforcement learning \citep{fisac2019bridging, hsu2021safety}, sum-of-squares (SOS) optimization \citep{landry2018reach} or 
differential dynamic programming (DDP)~\citep{pantoja1988differential}, in particular, using iterative linear-quadratic (ILQ) optimization~\citep{fridovichkeil2021approximate, nguyen2021back}.

\section{Reach-Avoid Control Barrier Functions}
While \new{least-restrictive} safety frameworks %
\new{provide maximum flexibility to task-driven control policies},
they may result in abrupt control switches.
\new{On the other hand, handcrafted CBFs are often overly conservative and difficult to design for high-dimensional systems.
All of the above methods rely on offline computation and suffer under unforeseen scenarios, such as obstacles emerging into the system's field of view.
}

\new{
This section introduces a novel scheme to systematically construct control barrier functions at runtime, by computing a reach-avoid value function through differential dynamic programming.
Specifically, we solve a \textit{receding-horizon} ILQ optimization problem until convergence and 
use the resulting value function to construct a quadratically constrained quadratic program (QCQP), allowing for a smooth transition between task-oriented and safety-oriented control.
We prove that the proposed CBF-DDP ensures infinite-horizon safety through recursively feasible finite-horizon planning by reaching a controlled-invariant target set.  
}

\out{This section introduces a technique that utilizes the linear quadratic (LQ) approximation of reach-avoid analysis to develop an implicit control barrier function.
Particularly, we solve ILQ optimization until convergence to obtain the reach-avoid value function $\valFunc$.
This function establishes a quadratically constrained quadratic program (QCQP), allowing for a smooth shift between task-oriented and safety-oriented control.
Unlike handcrafted CBFs, our CBF-DDP is general and can be constructed online.
This section outlines how to generate the barrier in real-time through a \textit{receding-horizon} approach to ensure infinite-horizon recursive feasibility through finite-horizon planning by reaching a controlled-invariant target set.}

\subsection{Barrier Safety Filter with Reach-Avoid Value Function}
\out{In this paper, we construct an implicit control barrier function by computing the LQ approximation of the HJ reach-avoid value function~\eqref{eq:hjb_reachavoid}. The reach-avoid value function is constructed in real-time using ILQ optimization.}%
We begin by laying out the theoretical foundations of our approach.
\new{In order to provide infinite-horizon safety guarantees, we assume the existence of a known target set with the following two properties:}
(1) the target set does not intersect with the failure set, $\target\cap\failure=\emptyset$, and 
(2) the target set is controlled-invariant under a known policy $\policy^\target$.

\new{
We note that the above are mild assumptions, satisfied by broad classes of systems of practical interest:
for example, many robots and vehicles are able to come to a stop, or alternatively transition into a safe loiter orbit, after which the system can remain safe indefinitely under benign engineering assumptions.
While more sophisticated methods may be used to determine the controlled-invariant regions of attraction around these states (e.g. Lyapunov analysis),
here we illustrate a simpler approach.
We construct such a target set using a conservative stopping policy $\policyStop$, which brings the robot to a complete halt and maintains the robot at rest.
The target margin function $\targFunc$ can be rapidly computed from each state as the signed distance to the failure set along the robot's hypothetical stopping path from that state. Being in the target set, then, is equivalent to being able to safely come to a halt using $\policyStop$.
Specifically, $\targFunc$ is defined as:
\begin{equation*}
    \targFunc\left( \state_\tdisc \right) = \min_{\tdiscaux \geq \tdisc} \consFunc \left(
        \traj \left( \tdiscaux-\tdisc; \state_\tdisc, \policyStop \right)
    \right).
    \vspace{-3mm}
\end{equation*}
}

\new{At each state $x_t$, we are provided with an arbitrary task-oriented control $\ctrl^\task$ proposed by the task policy $\policyTask$. For the corresponding control cycle $\tdisc$, we formulate a constrained program, which seeks to minimize deviation $\ctrlErr{}$ from $\ctrl^\task$ while satisfying a control barrier constraint as shown below:
\begin{subequations}
\begin{IEEEeqnarray}{cc}
    \min_{\ctrlErr{} \colon \ctrlErr{} + \ctrlTask \in \cSet} & \frac{1}{2}\left\| \ctrlErr{} \right\|_2^2  \\
    \such & \ \valFunc(\stateImag{1}{\tdisc}, 0) \geq \gamma \valFunc(\state_\tdisc, 0),~
    \label{eq:ddp_cbf_cstr}
\end{IEEEeqnarray}
\label{eq:ddp_cbf}%
\end{subequations}
where $\stateImag{1}{\tdisc} = \dyn(\state_\tdisc, \ctrl^\task+\ctrlErr{})$, $\valFunc(\state_\tdisc, 0)$ is the reach-avoid value~\eqref{eq:hjb_reachavoid} at the current state and $\valFunc(\stateImag{1}{\tdisc}, 0)$ is the reach-avoid value at the state reached after applying control  $\ctrlTask+\ctrlErr{}$.
Generally, computing these two values is computationally
prohibitive  %
for real-time decision-making.
Instead, we use DDP to compute the quadratic approximation of the reach-avoid value function and the (locally) optimal safe control.}
\out{
This LQ approximation provides a principled way to construct a CBF-type constraint.
Additionally, we apply this constraint in a receding-horizon fashion. We compute the LQ approximation for two value functions at each time step, denoted as $\valFunc^{0}$ and $\valFunc^{1}$.
Specifically, $\valFunc^{0}$ optimizes the control sequence locally to maximize the outcome in~\eqref{eq:outcome}.
In contrast, $\valFunc^{1}$ optimizes the same objective but fixes the first control to the arbitrary task-oriented control $\ctrl^\task$.}
\out{The ILQ gives us not only the (locally) optimal safe control (from $\valFunc^{0}$) but also a quadratic approximation of the reach-avoid value function after we take the task-oriented control (from $\valFunc^{1}$). At each time step $\tdisc$, we denote $\stateImag{0}{\tdisc} = \state_\tdisc$, $\ctrl^\task = \policyTask(\state_\tdisc)$, $\stateNom{1} = \dyn(\state_\tdisc, \ctrl^\task)$, and $\stateImag{1}{\tdisc} = \dyn(\state_\tdisc, \ctrl^\task+\ctrlErr{})$. We have}

\new{
We use the iterative linear-quadratic approach outlined in Alg.~\ref{alg:RALQ} to compute these value functions\out{due to its light-weight computation}. Additionally, $\valFunc(\stateImag{1}{\tdisc}, 0)$ in~\eqref{eq:ddp_cbf_cstr} is approximated by
}
\begin{align}
    \valFunc(\stateImag{1}{\tdisc}, 0) \approx \valFunc(\stateNom{1}, 0) + \stateErr{1}^\top \vx{} +  
    \frac{1}{2} \stateErr{1}^\top \vxx{} \stateErr{1},
\end{align}
\new{
where $\stateNom{1} = \dyn(\state_\tdisc, \ctrlTask)$ and $\stateErr{1} = \stateImag{1}{\tdisc} - \stateNom{1} \approx \fut{} \ctrlErr{}$ is obtained from dynamics Jacobian $\fut{} := \frac{\partial \dyn}{\partial \ctrl} |_{\state_\tdisc, \ctrlTask}$.
Then,~\eqref{eq:ddp_cbf_cstr} is approximated as a quadratic constraint:
}
\new{
${\ctrlErr{}^{\top}\cbfquad\ctrlErr{}+ \cbfaff^{\top} \ctrlErr{} + \cbfconst \geq 0}$, where ${\cbfquad = \frac{1}{2}\fut{}^{\top} \vxx{}\fut{}}$ is symmetric, $\cbfaff = \fut{}^{\top} \vx{}$, and $\cbfconst = \valFunc(\stateNom{1}, 0) - \gamma \valFunc(\state_\tdisc, 0)$. Thus,~\eqref{eq:ddp_cbf} is reduced to a quadratically constrained quadratic program (QCQP) as shown below:
\begin{subequations}
\begin{IEEEeqnarray}{cc}
    \min_{\ctrlErr{} \colon \ctrlErr{} + \ctrlTask \in \cSet} & \left\| \ctrlErr{} \right\|_2^2  \\
    \such & \ \quad \ctrlErr{}^{\top}\cbfquad\ctrlErr{}+ \cbfaff^{\top} \ctrlErr{} + \cbfconst \geq 0. ~
    \label{eq:ddp_cbf_qcqp_cstr}
\end{IEEEeqnarray}
\label{eq:ddp_cbf_qcqp}%
\end{subequations}
}
The final executed control is $\ctrl_\tdisc = \policyShield(\state_\tdisc) := \ctrl^\task + \ctrlErr{}^{\ast}$, where $\ctrlErr{}^{\ast}$ is an optimal solution to~\eqref{eq:ddp_cbf}. Henceforth, we represent the parameters of the QCQP in~\eqref{eq:ddp_cbf_qcqp} as $\optparam=(\cbfquad, \cbfaff, \cbfconst)$.

\begin{lemma}\label{lem:qcqp}
For the QCQP \new{with parameters $\optparam := (\cbfquad, \cbfaff, \cbfconst)$} in~\eqref{eq:ddp_cbf_qcqp}, if $\cbfquad$ is negative definite ($\cbfquad \prec 0$), then optimal solution $\ctrlErr{}^{\ast}$ is unique and continuous in $\optparam$. Additionally, if $\cbfquad = 0$, then $\ctrlErr{}^{\ast}$ is unique and Lipschitz continuous in $\optparam$.
\end{lemma}
\proof
If $\cbfquad=0$, the CBF-QP has a known closed-form solution. Thm.~3 in~\cite{ames2017cbf} proves Lipschitz continuity. If $\cbfquad \prec 0$, the QCQP is a strictly convex QP with a strictly convex constraint. Thus, $\ctrlErr{}^{\ast}$ is unique. The level sets of the constraint are bounded ellipsoids and are hence compact sets. The objective and quadratic constraint of~\eqref{eq:ddp_cbf_qcqp} are continuous in $\ctrlErr, \optparam$. From Berge's maximum theorem~\cite{berge1997topological}, the set of optimizers is compact and upper hemi-continuous (UHC). Since $\ctrlErr{}^{\ast}$ is unique, UHC is equivalent to continuity. 
\endproof

\begin{lemma}\label{lem:continuity}
    \final{Given $\forall \state, \ctrl$, almost everywhere (a.e.), $\valFunc \in \C^{2}$ with respect to (w.r.t.) $\state$ and $\dyn \in \C^{2}$ w.r.t. $(\state, \ctrl)$, $\policyTask$ is continuous in $\state$ and $\cbfquad\preceq 0$, then $\exists$ unique $\ctrlErr{}^{\ast}$ continuous in $\state$ a.e..} 
\end{lemma}
\proof
    Given $\valFunc, \dyn$ is $\C^{2}$ a.e., $\valFunc, \dyn$ is $\C^{2}$ at $\dyn(\state, \policyTask(\state))$ almost surely. Hence, $\optparam := (\cbfquad, \cbfaff, \cbfconst)$ is continuous in  $\state$ a.e.. Using Lemma~\ref{lem:qcqp} and the continuity of composition of continuous functions, $\ctrlErr{}^{\ast}$ is unique, continuous in $\state$ a.e.. 
\endproof
  A sufficient condition for continuous $\ctrlErr{}^{\ast}$ is if $\valFunc, \dyn$ is $\C^{2}$. We choose $\targFunc, \consFunc, \dyn$ to be piecewise $\C^{2}$ to aid this requirement. 
  To use Lemma \ref{lem:qcqp}, we check whether $\cbfquad \prec 0$ and switch to solving with a linear constraint if this condition is violated.
In practice, $\targFunc$ and $\consFunc$ are chosen as the maximum of multiple safety constraints that are independently $\C^{2}$. There can be isolated points of discontinuity in $\ctrl$. Compared to the least-restrictive baseline, these discontinuities appears away from the boundary of $\safeSetOpt$ and does not result in chatter along the boundary.
\begin{lemma}\label{lem:ideal_feasibility}
    Suppose $\state_\tdisc \in \raHoriFunc{\target}{\failure}{\khorizon}$ and $\state_\tdisc \notin \target$, then constraint \eqref{eq:ddp_cbf_cstr} is feasible. A feasible solution is $\ctrlErr{}=\ctrlBackup-\ctrlTask$ with
        $\ctrl^\backup=\arg \max_{\ctrl \in \cSet} \valFunc(\dyn(\state_\tdisc, \ctrl), 1).$
\end{lemma}
\proof   
\new{Since $\state_\tdisc \in \raHoriFunc{\target}{\failure}{\khorizon}$, we have $\valFunc(\state_\tdisc, 0) \geq 0$. Also, since $\state_\tdisc \notin \target$, $\targFunc(\state_\tdisc)<0$. Consequently, from \eqref{eq:hjb_reachavoid},    
\begin{align*}
        \valFunc(\state_\tdisc, 0)
            = \min \big\{& 
                \consFunc(\state_\tdisc), \\
                &\max_{\ctrl \in \cSet} \valFunc(\dyn(\state_\tdisc, \ctrl), 1)
            \big\}
            \leq \max_{\ctrl \in \cSet}  \valFunc(\dyn(\state_\tdisc, \ctrl), 1). \label{eq:observ1}
\end{align*}
Now, using \eqref{eq:outcome} and \eqref{eq:hjb_reachavoid}, for any $\state_0$,
    \begin{IEEEeqnarray*}{l}
        \valFunc\left( \state_0, 0 \right) \
        = \max_{\policyBackup} \max_{\tdiscaux \in [0, \khorizon]} \min \left\{
                \targFunc \big( \state_\tdiscaux \big),
                \min_{\tdiscauxaux \in [0, \tdiscaux]} \consFunc \big( \state_\tdiscauxaux \big)
            \right\}
        \nonumber \\
        \geq \max_{\policyBackup}
            \max_{\tdiscaux \in [0, \khorizon-1]} \min \Big\{
                \targFunc \big( \state_\tdiscaux \big),
                \min_{\tdiscauxaux \in [0, \tdiscaux]} \consFunc \big( \state_\tdiscauxaux \big)
            \Big\}
        = \valFunc \left( \state_0, 1 \right), \label{eq:observ2}
    \end{IEEEeqnarray*}
    where $\state_\tdiscaux = \traj(\tdiscaux; \state_0, \policyBackup)$.}
    From above observations, we note:
    \begin{align*}
        & \max_{\ctrl \in \cSet} \valFunc(\dyn(\state_\tdisc, \ctrl), 0)
            \geq \max_{\ctrl \in \cSet}  \valFunc(\dyn(\state_\tdisc, \ctrl), 1)
            > \gamma \valFunc(\state_\tdisc, 0).
    \end{align*}
    for any $\gamma \in \left[0, 1 \right)$.  This ensures that \out{$\exists u \in \cSet$} \new{$\exists \ctrlErr{}\colon \ctrlErr{} + \ctrlTask \in \cSet$} which guarantees the feasibility of~\eqref{eq:ddp_cbf_cstr}, e.g., \new{$\ctrlErr{}=\ctrlBackup-\ctrlTask$}. \out{with a feasible solution $\ctrl^\backup$}
\endproof
\noindent
With the above Lemmas, we can prove our results of interest.
\begin{theorem}[Recursive Safety]\label{theorem:recursive}
    Suppose that $\target\cap\failure=\emptyset$ and $\target$ is controlled-invariant. If $\state_\tdisc \in \raHoriFunc{\target}{\failure}{\khorizon}$, then $\exists \ctrl_\tdisc$ such that $\state_{\tdisc+1} = \dyn(\state_\tdisc, \ctrl_\tdisc) \in \raHoriFunc{\target}{\failure}{\khorizon}$. Furthermore, $\ctrl^\backup$ is a feasible solution for $\state_{\tdisc+1} \in \raHoriFunc{\target}{\failure}{\khorizon}$. 
\end{theorem}
\proof
    If $\state_\tdisc \in \target$, the theorem is proven by choice of $\target$ being controlled-invariant with a known policy $\policy^{\target}$.
    By definition, $\ctrl^\backup$ achieves $\valFunc(\dyn(\state_\tdisc, \ctrl^\backup), 1) \ge \valFunc(\dyn(\state_\tdisc, \policy^\target(\state_\tdisc)), 1) \ge 0$.
    If $\state_\tdisc \notin \target$, feasibility of~\eqref{eq:ddp_cbf_cstr} \new{from Lemma~\ref{lem:ideal_feasibility}} ensures \new{${\valFunc(\state_{\tdisc+1}, 0) \geq \gamma \valFunc(\state_\tdisc, 0) \geq 0}$ and therefore $\state_{\tdisc+1} \in \raHoriFunc{\target}{\failure}{\khorizon}$.}
    Choosing $\policyShield(\state)=\ctrl^\backup$ is sufficient for feasibility of~\eqref{eq:ddp_cbf_cstr}. 
\endproof

\begin{theorem}[Infinite-Horizon Safety]
    \new{Given $\state_{0} \in \target$, Alg.~\ref{alg:cbfddp} ensures $\state_{\tdisc}\in \raHoriFunc{\target}{\failure}{\khorizon},\forall \tdisc$. Further, Alg.~\ref{alg:cbfddp} maintains a fallback policy $\policyBackup$ ensuring safety for infinite-horizon.}
\end{theorem}
\proof
    Suppose $\state_{\tdisc} \in \raHoriFunc{\target}{\failure}{\khorizon}$, then from Thm. ~\ref{theorem:recursive}, we get $\state_{\tdisc+1} \in \raHoriFunc{\target}{\failure}{\khorizon}.$
    Given $\state_{0}\in \target$, we see $\state_{\tdisc}\in \raHoriFunc{\target}{\failure}{\khorizon},\forall \tdisc$.
    Also, given, $\state_{\tdisc} \in \raHoriFunc{\target}{\failure}{\khorizon}$, $\exists \policyBackupSig$ that ensures $\traj(\tdiscaux-\tdisc; \state_\tdisc, \policyBackupSig) \in \target$ for some $\tdiscaux \in [\tdisc, \tdisc+\khorizon]$.
    Alg.~\ref{alg:cbfddp} maintains a sequence of fallback policies $\policyBackupSig$ constructed by Alg.~\ref{alg:fallback} capable of guiding the robot inside the target set.
    The construction of $\policyBackupSig$ is valid by choice of using closed-loop policies from DDP until entering $\target$ and $\policyInv$ thereafter.
\endproof

\begin{theorem}[Continuity of controls]
    \final{ Given $\forall \state, \ctrl$, a.e., $\valFunc \in \C^{2}$ w.r.t. $\state$, $\dyn \in \C^{2}$ w.r.t. $(\state, \ctrl)$ and $\policyTask$ is continuous in $\state$, then Alg. \ref{alg:cbfddp} provides  $\ctrl_{\tdisc}$ continuous in $\state_{\tdisc}$ a.e. }
\end{theorem}
\proof
     Alg.~\ref{alg:cbfddp} solves~\eqref{eq:ddp_cbf_qcqp} iteratively until~\eqref{eq:ddp_cbf_cstr} is satisfied or the computational budget is exhausted. In each QCQP iteration, $\ctrlErr{}^{\ast}$ is continuous in $\state$ a.e. from Lemma~\ref{lem:continuity}. If the QCQP is not feasible, the reduction to QP is guaranteed to be feasible with $\ctrlErr{}^{\ast}$ continuous in $\state_\tdisc$ a.e.. Given $\state_\tdisc$, $\ctrlImag{1}{\tdisc}=\policyTask(\state_\tdisc) + \ctrlErr{}^{\ast}$ is a continuous function of $\state_\tdisc$ a.e. The final $\ctrlErr{}^{\ast}$ after multiple QCQP iterations is continuous if at each QCQP iteration, $\valFunc,\dyn \in \C^{2}$ at $\dyn(\state_\tdisc, \ctrlImag{1}{\tdisc})$. Alg.~\ref{alg:cbfddp} almost surely does not encounter a discontinuity during the QCQP iterations.  Thus, Alg.~\ref{alg:cbfddp} provides $\ctrl_{\tdisc}$ continuous in $\state_{\tdisc}$ a.e.
\endproof
\subsection{Implementation Details} \label{sec:impl}
The flow of our CBF-DDP is outlined in Alg.~\ref{alg:cbfddp}\footnote{The sequence of fallback policies can be initialized with $\policyBackupIdx{0:\khorizon-1} \gets \policyInv$.}. Each time, \new{the core subprocedure of the algorithm} solves the QCQP in~\eqref{eq:ddp_cbf_qcqp}. Since this QCQP \new{only} solves a local quadratic approximation of the non-linear constraint in~\eqref{eq:ddp_cbf_cstr}\new{, it} necessitates repeatedly updating the initial control and re-solving the QCQP until~\eqref{eq:ddp_cbf_cstr} is satisfied. To reduce the number of QCQPs we solve, we use a scaling parameter $\dynScale \new{\in \reals}$ on $\cbfconst$. With $\cbfconst<0$, by scaling $\cbfconst$ higher than needed, we get closer to satisfying~\eqref{eq:ddp_cbf_qcqp} faster. This inner loop is allowed to find a feasible solution within five attempts or is terminated. If QCQP cannot find a feasible solution and \out{$\valFunc^{1}<0$} $\valFunc(\stateImag{1}{\tdisc}, 0) < 0$, we use $\policyBackup$, thus guaranteeing fully safe operation. \final{This fallback never occurs in the simulations as long as $\gamma$ is chosen large enough that $\valFunc$ does not decrease rapidly to zero.}

\begin{algorithm}
\footnotesize
\caption{CBF-DDP}
\begin{algorithmic}[1]
\Require{$\state_{0} \in \target$, fallback policies (from $\state_{0}$) $\policyBackupSig$, policy for target set $\policyInv$, task policy $\policyTask$, scaling $\dynScale \geq 1$, CBF discount $\gamma \in (0, 1)$ }
\For {$\tdisc = 0, \dots, \thorizon$}
    \State $\ctrl^\task \gets \policyTask(\state_{\tdisc}), \stateNom{1} \gets \dyn (\state_{\tdisc}, \ctrl^\task)$
    \State $\valFunc(\stateNom{1}, 0), \fut{}, \vxx{}, \vx{}, \trajNom, \csigNom, \clsig, \olsig \gets$ ReachAvoidILQ$( \stateNom{1} )$ 
    \State $\cbfquad \gets \fut{}^{\top} \vxx{}\fut{}, \; \cbfaff \gets \fut{}^{\top} \vx{}, \; \cbfconst = \valFunc(\stateNom{1}, 0) - \gamma \valFunc(\state_\tdisc, 0)$
    \While{Equation \eqref{eq:ddp_cbf_cstr} \text{violated} \textbf{or not} \text{timeout}}
        \State $\ctrl^\task \gets$ BarrierFilter$( \cbfquad, \cbfaff, \cbfconst ), \stateNom{1} \gets \dyn (\state_{\tdisc}, \ctrl^\task)$\Comment{Solve QCQP~\eqref{eq:ddp_cbf}}
        \State $\valFunc(\stateNom{1}, 0), \fut{}, \vxx{}, \vx{}, \trajNom, \csigNom, \clsig, \olsig \gets$ ReachAvoidILQ$ (\stateNom{1})$ 
        \State $\cbfquad \gets \fut{}^{\top} \vxx{}\fut{},\; \cbfaff \gets \fut{}^{\top} \vx{}, \; \cbfconst = \valFunc(\stateNom{1}, 0) - \gamma \valFunc(\state_\tdisc, 0)$
        \State $\cbfconst \gets \dynScale \cbfconst$
    \EndWhile
    \If{$\valFunc(\stateNom{1}, 0)<0$}
        \State $\ctrl^\backup \gets \policyBackupIdx{0}(\state_\tdisc)$, $\policyBackupIdx{0:\khorizon - 2} \gets \policyBackupIdx{1:\khorizon - 1}$, $\policyBackupIdx{\khorizon - 1} \gets \policyInv$.
        \State Apply $\ctrlBackup$ \Comment{Fall back to full safety filter if $\ctrl^\task$ is unsafe}
    \Else
        \State $\policyBackupSig \gets$ Fallback$( \trajNom, \csigNom, \clsig, \olsig, \policyInv, \khorizon )$  \Comment{Update fallback policies}
        \State Apply $\ctrlTask$ \Comment{Apply control from CBF-DDP}
    \EndIf
\EndFor
	\end{algorithmic}\label{alg:cbfddp}
\end{algorithm}

\begin{algorithm}
\footnotesize
\caption{Reach-Avoid ILQ}
    \begin{algorithmic}[1]
    \Require{$\state_{0}$}
    \State Initialize \new{control sequence $\csigNom \gets 0$} (or warm-start with previous iteration), feedback gains $\clsig \gets 0$, open-loop gains $\olsig \gets 0$,
    and roll out for $\trajNom$
    \While{$\text{not converged}$}
        \LineComment{Find nominal trajectory, safety margins, and target margins}
        \State $\valFunc, \fut{}, \trajNom, \csigNom, \{ \targFunc_\tdisc, \consFunc_\tdisc \} \gets$
            Forward$(\trajNom, \csigNom, \clsig, \olsig )$
        \LineComment{Find local policies and a quadratic approximation of value function}
        \State $\{\closedloop, \openloop\}, \vxx{0}, \vx{0} \gets$
            Backward$(\trajNom, \csigNom, \{ \targFunc_\tdisc, \consFunc_\tdisc \} )$
    \EndWhile
    \State \Return $\valFunc, \fut{}, \vxx{0}, \vx{0}, \trajNom, \csigNom, \clsig, \olsig$
	\end{algorithmic}\label{alg:RALQ}
\end{algorithm}

\begin{algorithm}
\footnotesize
\caption{Construct Fallback Policy}
    \begin{algorithmic}[1]
        \Require $\trajNom, \csigNom, \clsig, \olsig, \policyInv, \khorizon$.
        \For{$\tdisc \gets 0$ to $\khorizon - 1$}
            \If{$\overline{\state}_\tdisc \in \target$}
                $\policyBackupIdx{\tdisc:\khorizon - 1} \gets \policyInv$ and break the for-loop
            \Else \ 
                $\policyBackupIdx{\tdisc}(\state_\tdisc) \gets \ctrlNom{\tdisc} + \closedloop (\state_\tdisc - \overline{\state}_\tdisc) + \openloop$
            \EndIf
        \EndFor
        \State \Return $\policyBackupSig$
    \end{algorithmic}\label{alg:fallback}
\end{algorithm}

\section{Simulation Results}
In many applications, such as autonomous driving, it is desirable to create a smooth safety filter that maintains safety desiderata. The least-restrictive safety filter causes jerky behavior, which affects the comfort of the rider in a car. We show our CBF-DDP method to result in a smoother and more desirable behavior. Our CBF-DDP algorithm uses $\dynScale \in [1.2, 1.3]$ in all our runs. In all simulations, we use the Runge–Kutta method (RK4) for integration with sampling time $\delta t=0.05$.
\subsection{Dubins Car}
We illustrate CBF-DDP in a Dubins car with velocity fixed at $0.7$. The equations of motion are
${\dot{x} = v\cos(\theta),\; \dot{y} = v\sin(\theta),\; \dot{\theta}=u_{1}}.$ A circular footprint with a radius of $0.15$ is chosen for the ego-vehicle. The steering control limits are $[-1.0, 1.0]$. We use a test case where the Dubins car goes around an obstacle. The task-oriented policy is to turn in the direction facing the goal. In this setup, we don't use a reach-avoid objective function. We use $\consFunc$ in a \textit{reachability} \cite{fridovichkeil2021approximate} formulation where the car is constrained to be in a set from which it avoids all obstacles.

\begin{figure}[!t]
    \centering
    \includegraphics[width=0.63\columnwidth]{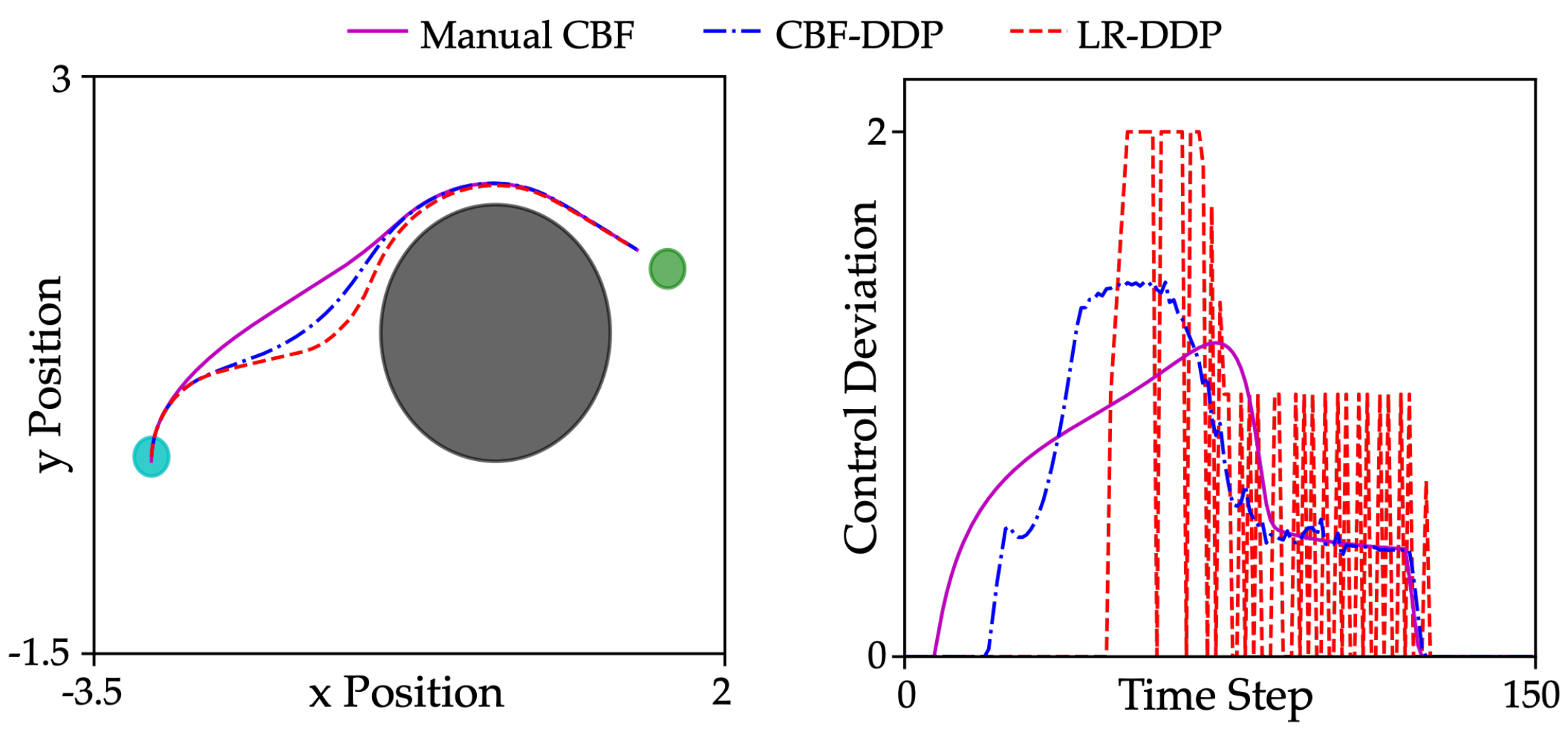}
    \caption{Dubins car: Comparison of safety filters CBF-DDP, LR-DDP and a manual CBF. (Left) Trajectories of the car. (Right) Steer control deviation from $\policyTask$.}
    \vspace{-13pt}
    \label{fig:dubins}
\end{figure}

The comparison of CBF-DDP with LR-DDP is illustrated in Fig. \ref{fig:dubins} with $\khorizon=40$. We see that LR-DDP filter \new{frequently switches between the task and safety control near the unsafe boundary, resulting in jerky movements}. In contrast, the CBF-DDP filter activates earlier and smoothly controls the car.
The number of QCQP iterations is limited to 2. We also compare it to a manual CBF that is known to be sub-optimal. The manual CBF provides smooth controls but is more conservative because of the sub-optimality.

\subsection{Kinematic Bicycle Dynamics}
We consequently test on dynamics with two controls, i.e., acceleration and steering. The equations of motion are
${\dot{x} = v\cos(\theta),\; \dot{y} = v\sin(\theta),\; \dot{v}=u_{1},\; \dot{\theta}=\frac{v}{L}\tan{\delta},\; \dot{\delta}=u_{2}}$.
\out{The car can slow down and curve around to avoid obstacles.} A circular footprint with a radius of $0.1$ is chosen for ego-vehicle. The acceleration control limits $u_{1}$ are $[-0.5, 0.5]$ and steering control limits $u_{2}$ are $[-0.8, 0.8]$.
Here, we use a reach-avoid value function with the target margin function computed based on the stopping path criterion.
\out{as illustrated in Fig. \ref{fig:cover_fig}
For each state, the target margin function $\targFunc$ is the distance to the failure set boundary along the path where the car is applying full deceleration to come to a halt. By lieu of this $\targFunc$, being in the reach-avoid set is equivalent to being able to come to a safe halt resulting in infinite horizon safety.}
For $\policyTask$, we use a naive heuristic where the steering is  pursuing a look-ahead point near the center of the road. This policy has a switch when the car is already moving in a desired direction. Further, this $\policyTask$ feeds acceleration to maintain velocity at $0.9$ using linear feedback. \new{We run our experiments on a 3.6 GHz CPU with 64 GB RAM.}

\begin{figure*}[!t]
    \centering
    \includegraphics[width=0.85\textwidth]{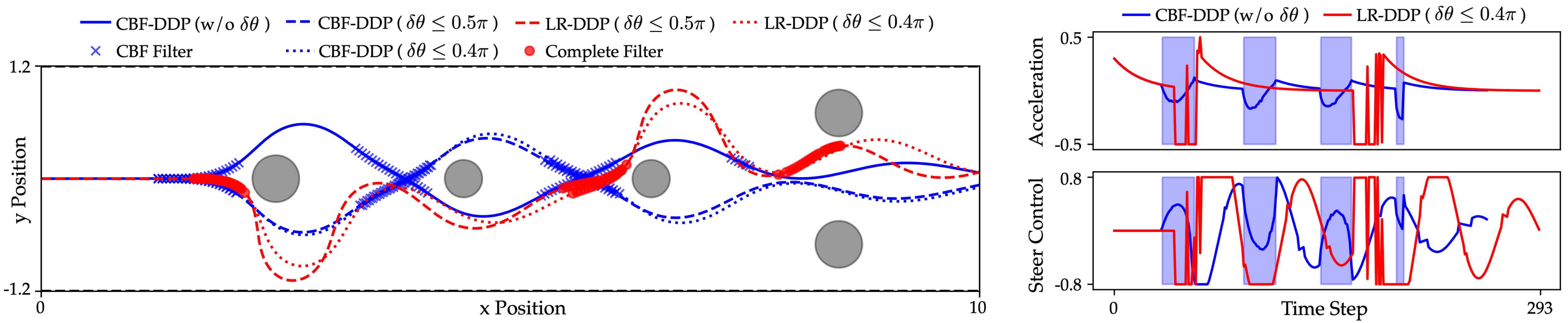}
    \caption{(Left) Comparison of CBF-DDP and LR-DDP with varying $\yawdeviation$. Trajectories from CBF-DDP are much more well-behaved. Videos available at \url{https://tinyurl.com/2pr8dy6j}. (Right) Comparison of controls from CBF-DDP and LR-DDP with favorable $\yawdeviation$ for each. CBF-DDP gives smoother controls in the blue-shaded regions where the filter is active.}
    \label{fig:bic}
    \vspace{-9pt}
\end{figure*}

The CBF-DDP is robust and provides smooth controls (Fig. \ref{fig:bic}) for these dynamics, but LR-DDP  makes a U-turn or comes to a safe halt resulting in task failure. 
LR-DDP can be improved by constraining the horizontal yaw deviation $\yawdeviation$. The $\targFunc, \consFunc$ are then defined using the worst-case margin of all constraints. 
The CBF-DDP provides smoother trajectories and fast task completion (Fig. \ref{fig:bic}) even with the constraint, but the controls get jerkier with tighter
limits on $\yawdeviation$.  \new{Discontinuities in derivatives of $\valFunc$ from interacting constraints between two obstacles or between obstacle and yaw can induce discontinuities in controls from CBF-DDP. Future practical improvements can incorporate soft versions of minimum operators to ease some of these discontinuities.}  Table~\ref{tab:yaw_longt_config_quant}
compares LR-DDP and CBF-DDP with \new{$\khorizon=45$} and a constraint on $\yawdeviation$. LR-DDP completes the task, but the trajectories are quantitatively jerkier than CBF-DDP. From Table~\ref{tab:yaw_longt_config_quant}, the total integral deviation ($\Vert \policyTask - \policyShield \Vert_{1}$) from $\policyTask$ is less for CBF-DDP, indicating that it is not overly restrictive. With stronger task policies, the yaw constraint is not needed giving an advantage to CBF-DDP.
\out{For $\policyTask$, we 
also 
evaluate using receding horizon ILQ optimization with the sum of multiple cost functions. This cost penalizes soft constraint violations such as yaw, velocity, and track. %
The additive weight of each soft constraint is tuned such that $\policyTask$ maintains the vehicle near the center of the road at a reference velocity of $0.9$. With this much-improved task policy, the yaw constraint is unnecessary for LR-DDP to complete the task (Fig. \ref{fig:bic}). The qualitative behavior of CBF-DDP compared to LR-DDP remains unchanged (Table \ref{tab:yaw_longt_config_quant}), demonstrating the benefits of CBF-DDP.}

\begin{table*}
\centering
\scriptsize
\begin{tabular}     {|C{1.1cm}|C{0.8cm}|C{0.4cm}|C{0.4cm}|C{1.3cm}|C{1.4cm}|C{1.3cm}|C{1.3cm}|C{1cm}|C{1cm}|C{1.3cm}|C{1.3cm}|}
    \hline
    \multicolumn{2}{|c|}{\centering Constraints} &
    \multicolumn{2}{c|}{Task Success} & 
    \multicolumn{2}{c|}{Acceleration Jerk (mean and std) } & 
    \multicolumn{2}{c|}{Steer Jerk (mean and std)} & 
    \multicolumn{2}{c|}{{Total Deviation from \new{$\policyTask$}}} &
    \multicolumn{2}{c|}{Control cycle time (s)}
    \\
    \cline{1-12}
    Road limit & yaw $\yawdeviation$ & LR & CBF & LR & CBF & LR & CBF & LR & CBF & LR & CBF \\
    \hline
    1.2& None & $\times$ & $\checkmark$ & $\times$ & 0.009 $\pm $0.02 & $\times$ & 0.05 $\pm$ 0.06 & $\times$ & 52.81 & $\times$ & 0.10 $\pm$ 0.10\\
    \hline
    1.2& $0.5\pi$  & $\checkmark$ &  $\checkmark$ & 0.03 $\pm$ 0.14 & 0.01 $\pm$ 0.02 & 0.06 $\pm$ 0.19 & 0.05 $\pm$ 0.08 & 93.23 & 50.46 & 0.21 $\pm$ 0.22 & 0.27 $\pm$ 0.39 \\
    \hline
    1.2& $0.4\pi$  & $\checkmark$ &  $\checkmark$ & 0.03 $\pm$ 0.15 & 0.01 $\pm$ 0.04 & 0.10 $\pm$ 0.27 & 0.08 $\pm$ 0.13 & 59.50 & 53.79 &  0.22 $\pm$ 0.20 & 0.38 $\pm$ 0.52\\
    \hline
    1.0& None & $\times$  & $\checkmark$  & $\times$ & 0.009 $\pm$ 0.02  & $\times$& 0.04 $\pm$ 0.06 & $\times$ & 52.81 & $\times$& 0.10 $\pm$ 0.09\\
    \hline
    1.0& $0.5\pi$  & $\times$ & $\checkmark$ & $\times$ & 0.01 $\pm$ 0.03 & $\times$ & 0.06 $\pm$ 0.11 & $\times$& 49.47 & $\times$& 0.28 $\pm$ 0.39\\
    \hline
    1.0& $0.4\pi$& $\checkmark$ &  $\checkmark$ &  0.05 $\pm$ 0.17 & 0.01 $\pm$ 0.03 &  0.11 $\pm$ 0.32 & 0.07 $\pm$ 0.11 &  78.15 & 51.93 & 0.20 $\pm$ 0.19 & 0.37 $\pm$ 0.52\\
    \hline
\end{tabular}
\caption{Bicycle: Comparison of LR-DDP vs CBF-DDP in obstacle avoidance with naive $\policyTask$. The maximum allowable yaw deviation from the horizontal $\yawdeviation$ is varied. CBF-DDP is robust to road boundaries and yaw constraints. The discontinuity of controls in CBF-DDP increases as the yaw constraint is tightened. On the other hand, LR-DDP tends to halt or U-turn safely without a yaw constraint.}\label{tab:yaw_longt_config_quant}
\end{table*}

\section{Conclusions}
\looseness=-1
We propose a principled approach to construct a CBF online with reach-avoid DDP.
We use a receding horizon to provide infinite-horizon recursive feasibility despite finite-horizon planning. The construction of our reach-avoid value functions guarantees safety.
Simulations  on a bicycle model show that CBF-DDP improves the smoothness over least-restrictive control and avoids the conservativeness of handcrafted CBFs. 
\out{Compared to the least-restrictive baseline, the safety filter acts much earlier when approaching the unsafe set boundary and provides faster and smoother obstacle avoidance.} 

\balance
\printbibliography

@inproceedings{hsu2021safety,
  author    = {Kai-Chieh Hsu and Vicenç Rubies-Royo and Claire J. Tomlin and Jaime F. Fisac},
  title     = {Safety and Liveness Guarantees through Reach-Avoid Reinforcement Learning},
  booktitle = {Proceedings of Robotics: Science and Systems},
  year      = {2021},
  address   = {Virtual},
  month     = {7},
  doi       = {10.15607/RSS.2021.XVII.077}
}

@misc{bui2022optimizeddp,
  doi = {10.48550/ARXIV.2204.05520},
  url = {https://arxiv.org/abs/2204.05520},
  author = {Bui, Minh and Giovanis, George and Chen, Mo and Shriraman, Arrvindh},
  title = {{OptimizedDP}: An Efficient, User-friendly Library For Optimal Control and Dynamic Programming},
  publisher = {arXiv},
  year = {2022},
  copyright = {Creative Commons Attribution 4.0 International}
}

@article{ames2017cbf,
  author={Ames, Aaron D. and Xu, Xiangru and Grizzle, Jessy W. and Tabuada, Paulo},
  journal={IEEE Transactions on Automatic Control}, 
  title={Control Barrier Function Based Quadratic Programs for Safety Critical Systems}, 
  year={2017},
  volume={62},
  number={8},
  pages={3861-3876},
  doi={10.1109/TAC.2016.2638961}
}

@inproceedings{agrawal2017discrete, 
  AUTHOR    = {Ayush Agrawal AND Koushil Sreenath}, 
  TITLE     = {Discrete Control Barrier Functions for Safety-Critical Control of Discrete Systems with  Application to Bipedal Robot Navigation}, 
  BOOKTITLE = {Proceedings of Robotics: Science and Systems}, 
  YEAR      = {2017}, 
  ADDRESS   = {Cambridge, Massachusetts}, 
  MONTH     = {7}, 
  DOI       = {10.15607/RSS.2017.XIII.073} 
}

@inproceedings{choi2021robust,
  author={Choi, Jason J. and Lee, Donggun and Sreenath, Koushil and Tomlin, Claire J. and Herbert, Sylvia L.},
  booktitle={Proceedings of the 60th IEEE Conference on Decision and Control (CDC)}, 
  title={Robust Control Barrier-Value Functions for Safety-Critical Control}, 
  year={2021},
  volume={},
  number={},
  pages={6814-6821},
  doi={10.1109/CDC45484.2021.9683085}
}

@inproceedings{chen2021backup,
  author={Chen, Yuxiao and Jankovic, Mrdjan and Santillo, Mario and Ames, Aaron D.},
  booktitle={Proceedings of the 60th IEEE Conference on Decision and Control (CDC)}, 
  title={Backup Control Barrier Functions: Formulation and Comparative Study}, 
  year={2021},
  volume={},
  number={},
  pages={6835-6841},
  doi={10.1109/CDC45484.2021.9683111}
}

@article{mitchell2008flexible,
  title={The flexible, extensible and efficient toolbox of level set methods},
  author={Mitchell, Ian M},
  journal={Journal of Scientific Computing},
  volume={35},
  number={2},
  pages={300--329},
  year={2008},
  publisher={Springer},
  DOI = {10.1007/s10915-007-9174-4}
}

@inproceedings{bansal2017hamilton,
  author={Bansal, Somil and Chen, Mo and Herbert, Sylvia and Tomlin, Claire J.},
  booktitle={Proceedings of the IEEE Annual Conference on Decision and Control (CDC)}, 
  title={Hamilton-Jacobi reachability: A brief overview and recent advances}, 
  year={2017},
  volume={},
  number={},
  pages={2242-2253},
  doi={10.1109/CDC.2017.8263977}
}

@inproceedings{landry2018reach,
  author={Landry, Benoit and Chen, Mo and Hemley, Scott and Pavone, Marco},
  booktitle={Proceedings of the IEEE/RSJ International Conference on Intelligent Robots and Systems (IROS)}, 
  title={Reach-Avoid Problems via Sum-or-Squares Optimization and Dynamic Programming}, 
  year={2018},
  volume={},
  number={},
  pages={4325-4332},
  doi={10.1109/IROS.2018.8594078}
}

@article{chen2018decomposition,
  author={Chen, Mo and Herbert, Sylvia L. and Vashishtha, Mahesh S. and Bansal, Somil and Tomlin, Claire J.},
  journal={IEEE Transactions on Automatic Control}, 
  title={Decomposition of Reachable Sets and Tubes for a Class of Nonlinear Systems}, 
  year={2018},
  volume={63},
  number={11},
  pages={3675-3688},
  doi={10.1109/TAC.2018.2797194}
}

@inproceedings{fridovichkeil2021approximate,
  author={Fridovich-Keil, David and Tomlin, Claire J.},
  booktitle={Proceedings of the IEEE International Conference on Robotics and Automation (ICRA)},
  title={Approximate Solutions to a Class of Reachability Games},
  year={2021},
  volume={},
  number={},
  pages={12610-12617},
  doi={10.1109/ICRA48506.2021.9561655}
}

@inproceedings{fisac2019bridging,
  author={Fisac, Jaime F. and Lugovoy, Neil F. and Rubies-Royo, Vicenç and Ghosh, Shromona and Tomlin, Claire J.},
  booktitle={Proceedings of the International Conference on Robotics and Automation (ICRA)},
  title={Bridging Hamilton-Jacobi Safety Analysis and Reinforcement Learning},
  year={2019},
  volume={},
  number={},
  pages={8550-8556},
  doi={10.1109/ICRA.2019.8794107}
}

@misc{nguyen2021back,
  doi = {10.48550/ARXIV.2109.07673},
  author = {Anthony, Dennis R. and Nguyen, Duy P. and Fridovich-Keil, David and Fisac, Jaime F.},
  title = {Back to the Future: Efficient, Time-Consistent Solutions in Reach-Avoid Games},
  organization = {arXiv},
  year = {2021},
  copyright = {Creative Commons Attribution 4.0 International}
}

@article{pantoja1988differential,
    author = {J. F. A.   DE O. PANTOJA },
    title = {Differential dynamic programming and Newton's method},
    journal = {International Journal of Control},
    volume = {47},
    number = {5},
    pages = {1539-1553},
    year  = {1988},
    publisher = {Taylor & Francis},
    doi = {10.1080/00207178808906114}
}

@article{mayne2000constrained,
    title = {Constrained model predictive control: Stability and optimality},
    journal = {Automatica},
    volume = {36},
    number = {6},
    pages = {789-814},
    year = {2000},
    issn = {0005-1098},
    doi = {https://doi.org/10.1016/S0005-1098(99)00214-9},
    author = {D.Q. Mayne and J.B. Rawlings and C.V. Rao and P.O.M. Scokaert}
}

@inproceedings{fisac2015reach,
    author = {Fisac, Jaime F. and Chen, Mo and Tomlin, Claire J. and Sastry, S. Shankar},
    title = {{Reach-Avoid Problems with Time-Varying Dynamics, Targets and Constraints}},
    year = {2015},
    isbn = {9781450334334},
    address = {New York, NY, USA},
    doi = {10.1145/2728606.2728612},
    booktitle = {Proceedings of the International Conference on Hybrid Systems: Computation and Control},
    pages = {11-20},
    location = {Seattle, Washington},
}

@book{berge1997topological,
  title={Topological Spaces: including a treatment of multi-valued functions, vector spaces, and convexity},
  author={Berge, Claude},
  year={1997},
  publisher={Courier Corporation}
}

@INPROCEEDINGS{tonkens2022refining,
  author={Tonkens, Sander and Herbert, Sylvia},
  booktitle={Proceedings of the IEEE/RSJ International Conference on Intelligent Robots and Systems (IROS)}, 
  title={Refining Control Barrier Functions through Hamilton-Jacobi Reachability}, 
  year={2022},
  volume={},
  number={},
  pages={13355-13362},
  doi={10.1109/IROS47612.2022.9982203}
}

\end{document}